\documentclass{article}

\usepackage[numbers]{natbib}

\usepackage{arxiv}

\usepackage[utf8]{inputenc} 
\usepackage[T1]{fontenc}    
\usepackage{hyperref}       
\usepackage{url}            
\usepackage{booktabs}       
\usepackage{amsfonts}       
\usepackage{nicefrac}       
\usepackage{microtype}      
\usepackage{lipsum}

\usepackage{booktabs} 
\usepackage{lineno,hyperref}

\usepackage{todonotes}
\usepackage{makecell}
\usepackage{graphicx}
\usepackage{adjustbox}
\usepackage{paralist}
\usepackage{amsmath}
\usepackage{booktabs}
\usepackage{tabularx}
\usepackage{array}
\usepackage[nameinlink]{cleveref}
\usepackage{verbatim}
\usepackage{comment}

\modulolinenumbers[5]

\newcolumntype{R}[1]{>{\tiny\raggedleft\arraybackslash}b{#1}}

\usepackage[ruled]{algorithm2e} 

\title{An Empirical Survey on Co-simulation: Promising Standards, Challenges and Research Needs}

\author{
  Gerald Schweiger \\
  Graz University of Technology\\
  Graz, Austria \\
  \texttt{gerald.schweiger@tugraz.at} \\
   \And
 Claudio Gomes \\
  University of Antwerp\\
  Antwerp, Belgium \\
  \texttt{claudio.gomes@uantwerp.be} \\
     \And
 Georg Engel \\
  Graz University of Technology\\
  Graz, Austria \\
  \texttt{engel@ist.tugraz.at} \\
     \And
 Josef-Peter Schoeggl \\
  KTH Royal Institute of Technology\\
  Stockholm, Sweden \\
  \texttt{ schoggl@kth.se} \\
     \And
 Alfred Posch \\
  University of Graz\\
  Graz, Austria \\
  \texttt{alfred.posch@uni-graz.at} \\
     \And
           Irene Hafner \\
  dwh - Simulation Services und Technical Solutions\\
  Vienna, Austria \\
  \texttt{irene.hafner@dwh.at} \\
     \And
 Thierry Nouidu \\
  Lawrence Berkeley National Laboratory\\
  Berkeley, USA \\
  \texttt{tsnouidui@lbl.gov} \\
  }

\begin{document}
\maketitle

\begin{abstract}
Co-simulation is a promising approach for the modelling and simulation of complex systems, that makes use of mature simulation tools in the respective domains.
It has been applied in wildly different domains, oftentimes without a comprehensive study of the impact to the simulation results. As a consequence, over the recent years, researchers have set out to understand the essential challenges arising from the application of this technique.
This paper complements the existing surveys in that the social and empirical aspects were addressed.
More than 50 experts participated in a two-stage Delphi study to determine current challenges, research needs and promising standards and tools.
Furthermore, an analysis of the strengths, weakness, opportunities and threats of co-simulation utilizing the analytic hierarchy process resulting in a SWOT-AHP analysis is presented.
The empirical results of this study show that experts consider the FMI standard to be the most promising standard for continuous time, discrete event and hybrid co-simulation.
The results of the SWOT-AHP analysis indicate that factors related to strengths and opportunities predominate.
\end{abstract}

\section{Introduction}

Simulation-driven assessments and developments are key methods used in various fields in industry and academia such as energy systems, production industries and social sciences. 
Due to the increasing complexity of systems, market competition and specialization, evaluating the overall behavior of these systems at every stage of their development is becoming steadily more difficult, ranging from early \emph{what-if} architectural analyses to detailed three dimensional simulations.
In order to keep benefiting from the results of simulation-based analyses, new techniques are required to efficiently simulate the interactions between subsystems. 
There are two ways to achieve this end: 
\begin{inparaenum}[(i)]
\item the entire system can be modelled and simulated with a single tool which is referred to as monolithic simulation; or
\item established tools for the respective subsystems can be coupled in a so-called co-simulation. 
\end{inparaenum}

As our knowledge of each subsystem matures, simulation tools become more specialized, accumulating years of research and practical experience in their respective domains. 
As such, the use of the co-simulation approach allows existing simulation tools to be leveraged. 
It has the potential to provide a quick and accurate way to realize holistic simulations by depicting interactions between subsystems while using the most appropriate simulators for each subsystem \cite{VanderAuweraer2013}. 
In co-simulation, the subsystem models are interconnected at their behavioural levels, through the traces computed corresponding simulation tools. 
Each simulation tool is seen as a black box, capable of producing outputs and consuming inputs, according to the model it represents.
This approach is attractive because:
\begin{compactitem}
\item The trace level is the simplest level at which any subsystem integration can be performed, and any dynamic model can be interpreted to produce such traces;
\item Each black box incorporates its own simulation algorithm, which is usually the most appropriate for its domain;
\item The exchange of the black box models can be made without requiring their content to be disclosed, thereby protecting IP and avoiding licensing fees.
\end{compactitem}
As a consequence, there has been an increasing number of applications of co-simulation across many domains \cite{Gomes2017,Brembeck2011,Pedersen2017,Engel2018,Sanfilippo2018,Schweiger2018c}

Unfortunately, naively connecting inputs to outputs on black boxes does not necessarily imply that the resulting behavior mimics the actual couplings of the subsystem models, which leads to a the identification of a central research question in co-simulation: \emph{are the co-simulation results trustworthy?}
This is not a new question, and the coupling of simulators can be traced back to multi-rate simulation techniques \cite{Gomes2018b}. 
However, the black box nature of co-simulation makes it unique.
For this reason, researchers have begun to address this challenge in the different domains in which co-simulation is applied.

\begin{figure}[htb]
\centering
\includegraphics[width=1\textwidth]{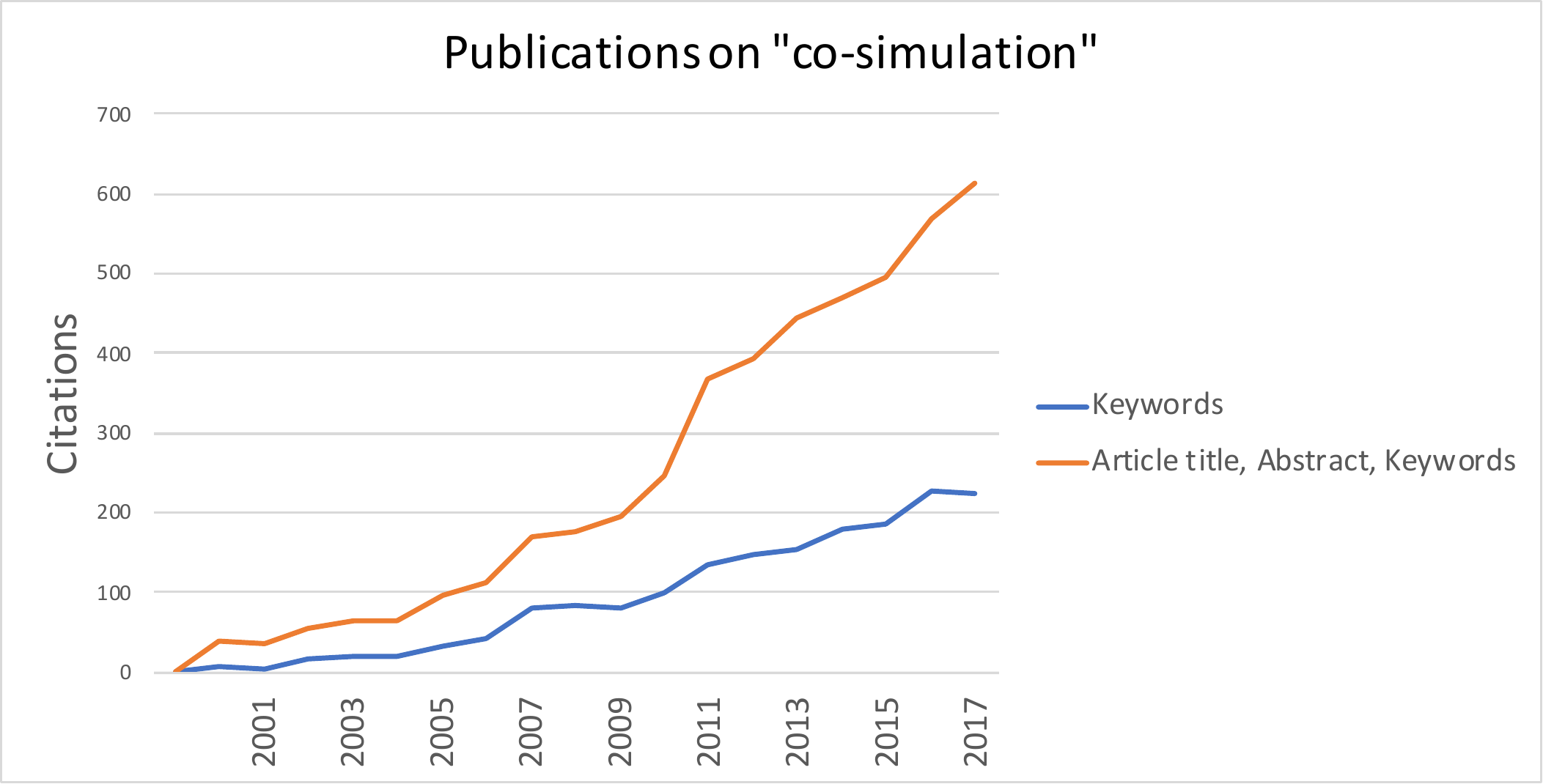}
\caption{Publications that included the keyword "co-simulation"}
\label{fig:Publication}
\end{figure}

\begin{figure}[htb]
\centering
\includegraphics[width=1\textwidth]{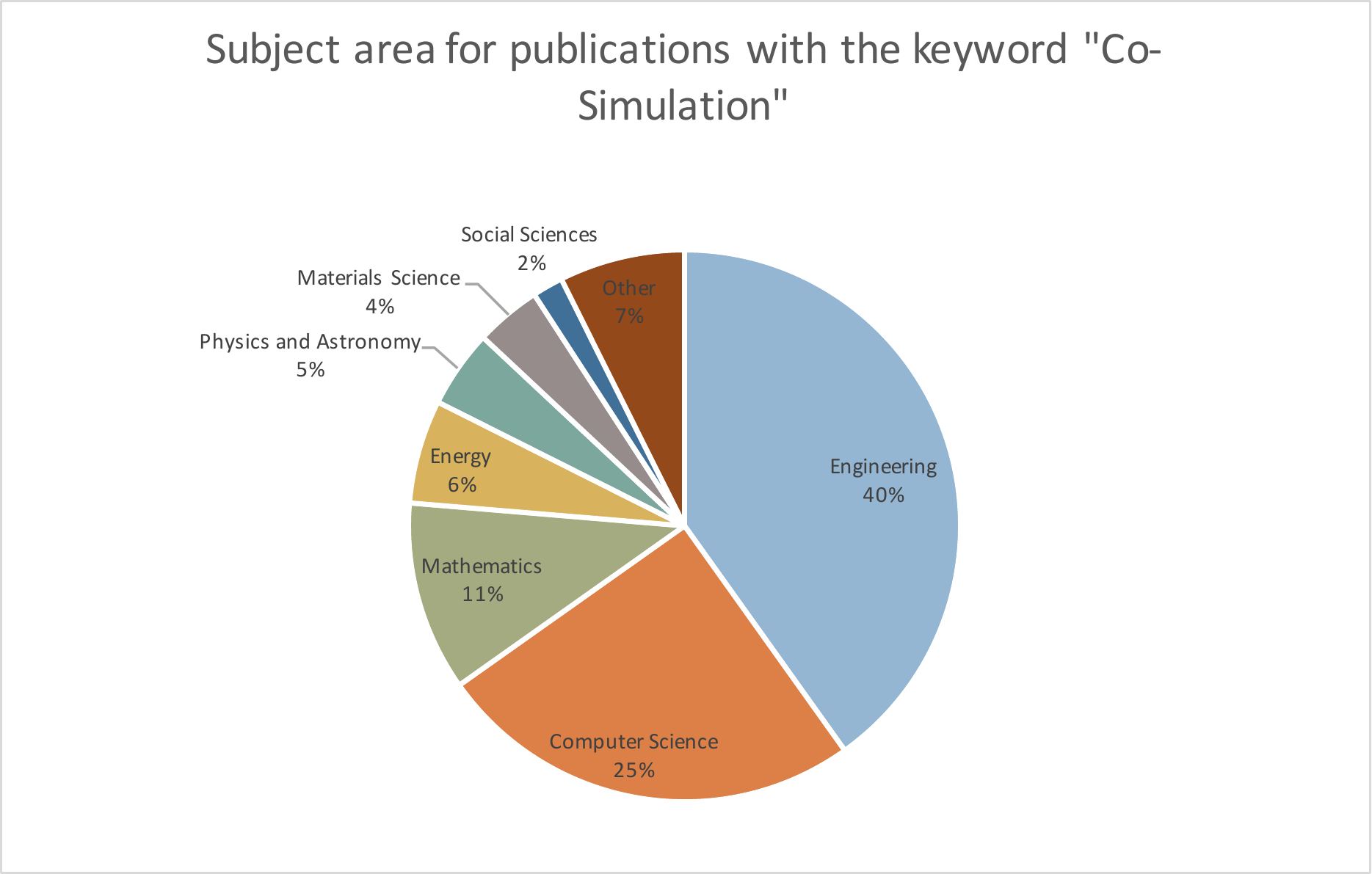}
\caption{Subject area for publications that include the keyword "co-simulation"}
\label{fig:Subject}
\end{figure}

To assess the importance of co-simulation in the scientific community, we conducted a keyword analysis and examined co-simulation related projects. 
The analysis was performed on Scopus with the keyword ``co-simulation''. 
\Cref{fig:Publication} shows that the number of citations grew in an almost linear fashion from $2000$ to $2017$. 
As it can be seen in \Cref{fig:Subject}, most of the publications can be assigned to the fields of Engineering ($40\%$) followed by Computer Science ($25\%$) and Mathematics ($11\%$).
\Cref{tab:projects} gives an overview of prominent recent research projects 
related to co-simulation.

\begin{table}[htb]
  \scriptsize
  \caption{Excerpt of research activities in the field of co-simulation in recent years (not complete)}
  \label{tab:projects}
  \begin{tabular}{llp{0.72\textwidth}}
  \textbf{Project} & \textbf{Duration} & \textbf{Goals} \\ \hline
  COSIBA~\cite{cosibas} & 2000--2002 &  Formulate a co-simulation backplane for coupling electronic design automation tools,  supporting different abstraction levels. \\
  ODETTE~\cite{odette} & 2000--20003 &  Develop a complete co-design solution including hardware/software co-simulation and synthesis tools. \\
  MODELISAR~\cite{modelisar} & 2008--2011 &  Improve the design of embedded software in vehicles. \\
  DESTECS~\cite{destecs} & 2010--2012 &  Improve the development of fault-tolerant embedded systems. \\
  INTO-CPS~\cite{intocps} & 2015--2017 &  Create an integrated tool chain for Model-Based Design of CPS with FMI.  \\
  ACOSAR~\cite{acosar} & 2015--2018 &  Develop a non-proprietary advanced co-simulation interface for real time system integration.  \\
  OpenCPS~\cite{opencps} & 2015--2018 &  Improve the interoperability between Modelica, UML and FMI.  \\
  ERIGrid~\cite{erigrid} & 2015--2020 &  Propose solutions for Cyber-Physical Energy Systems through co-simulation.  \\
  PEGASUS~\cite{pegasus} & 2016--2019 &  Establish standards for autonomous driving. \\
  CyDER~\cite{cyder} & 2017--2020 &  Develop a co-simulation platform for integration and analysis of high PV penetration. \\
  EMPHYSIS~\cite{emphysis} & 2017--2020 &  Develop a new standard (eFMI) for modeling and simulation environments of embedded systems.  \\ 
  \hline
  \end{tabular}%
\end{table}
The extensive and scattered nature of the body of knowledge in co-simulation, prompted researchers to initiate surveys of the state of the art.

\subsection{State of the Art}
\label{sec:related_work}

Co-simulation is now widely used in industry and academia.
This has motivated researcher to conduct survey work and examine fundamental concepts of co-simulation and the terminology used.
A discussion of differences in terminology and an attempt to classify and structure co-simulation methods was made by Hafner and Popper \cite{Hafner2017}.
The authors proposed several possibilities of classifying and structuring methods of co-simulation: (i) distinction by the State of Development, (ii) distinction by the field of application, (iii) distinction by the model description, (iv) distinction by numeric approaches and (v) distinction by Interfaces. 
Furthermore, a classification of multi-rate methods was proposed. 
Recognizing that co-simulation is not a new concept and that it has been applied in wildly different fields, Gomes et al. \cite{Gomes2018} reviewed co-simulation approaches, research challenges, and research opportunities. 
They applied a feature oriented domain analysis method \cite{Kang1990} to help map the field. 
The main result was a feature model that could be used to classify the requirements of co-simulation frameworks and the participating simulators. 
They concluded that the main research needs were: finding generic approaches for modular, stable, valid and accurate coupling of simulation units and finding standard interfaces for hybrid co-simulation. 
Trcka and Wetter \cite{Trcka2007} reviewed (i) principles and strategies of co-simulation including a discussion of the terminology, (ii) the topic of stability and accuracy within co-simulation, (iii) tools and communication mechanisms that are used in prototypes, and (iv) verification and validation techniques. 
Based on numerical experimentation and case studies they conclude that the advantages  of co-simulation are the flexibility by combining features from different tools; disadvantages were the difficulty of use and the required knowledge.
Placing a focus on power systems, but still covering the fundamental concepts, Palensky et al. \cite{Palensky2017} highlighted the value of co-simulation for the analysis of the former. 
In a tutorial fashion, they go over the main concepts and challenges, providing a great introduction for new researchers in the field.
To the best of our knowledge, these efforts are purely based on the existing literature and lack an empirical aspect.


\subsection{Main Contribution}

This work complements the existing surveys by providing the empirical aspect. 
We interviewed multiple experts from various fields in industry and academia as part of a two-stage Delphi study. 
As a result, the current challenges, research needs, and promising standards and tools were investigated using qualitative and quantitative research methods. 
Some of the challenges identified by the experts indeed match the conclusions of the existing surveys.
As such, the current work allowed us to rank the existing research according to their importance, as perceived by industry and academia.
Furthermore, we present an analysis of the internal strengths and weaknesses as well as the external opportunities for and threats to co-simulation by combining a SWOT (strengths, weaknesses, opportunities and threats) analysis with the AHP (Analytic Hierarchy Process) resulting in an SWOT-AHP analysis, the results of which allowed us to gain a better understanding of the relative importance of the respective factors.

The findings in the present work: 
\begin{compactitem}
  \item contribute to the structured and focused further development of various disciplines within the co-simulation community;
  \item can guide the efforts of the scientific community to address problems that are directly relevant to industry; and
  \item can serve as a practical guide by providing references to existing surveys, promising standards and tools  for co-simulation.
\end{compactitem}

A detailed discussion of the technical challenges goes beyond the scope of this work. Instead, we will provide the relevant references when appropriate.

The paper is structured as follows: In Chapter 2, we present the background including a discussion of different co-simulation approaches. In Chapter 3, we present a detailed presentation of the proposed method. Chapter 4 provides the results and discussion of the empirical survey and Chapter 5, the conclusion of the study.

\section{Background on Co-simulation}



\begin{figure}[tbh]
\begin{center}
  \includegraphics[width=0.9\textwidth]{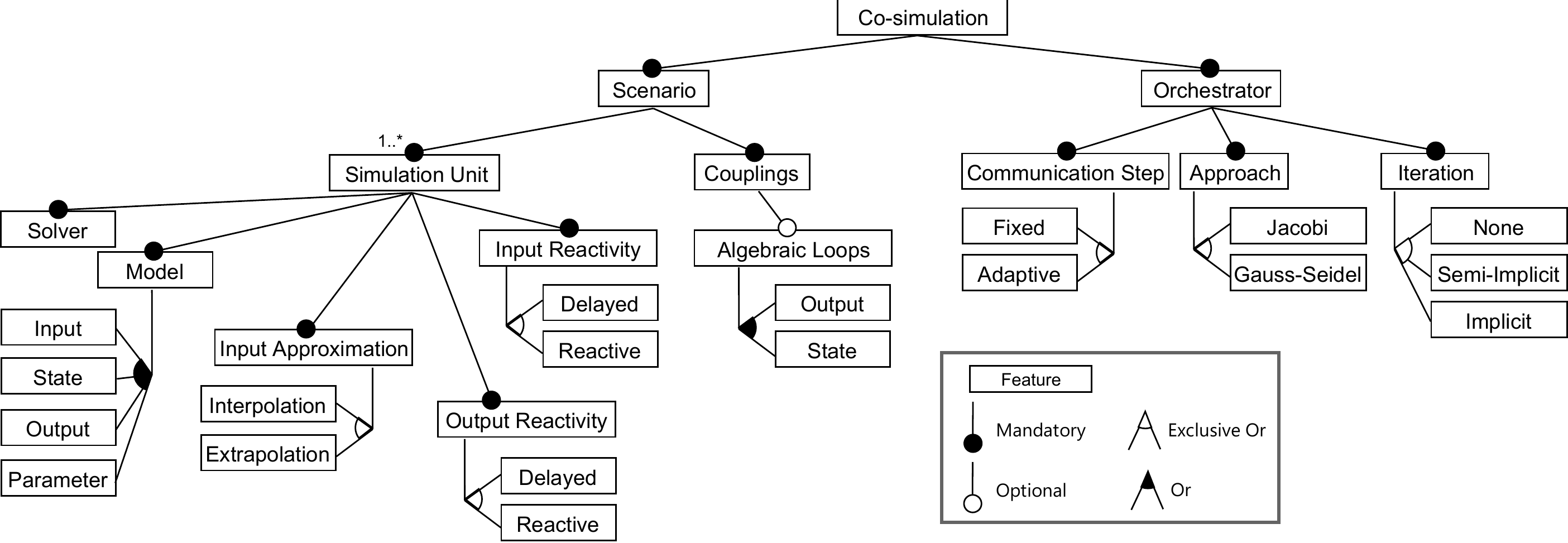}
  \caption{Co-simulation concept breakdown. Taken from \cite{Gomes2018e}.}
  \label{fig:concept_breakdown}
\end{center}
\end{figure}

In this section, we will provide an informal, top-down overview on the concepts related to co-simulation, based on \cite{Gomes2018e}.
To that end, we will use a \emph{feature model} \cite{Kang1990}: an intuitive diagram that breaks down the main concepts in a domain.
More rigorous definitions are given in \cite{Gomes2018}.
First, we summarize the \emph{objective} of running a co-simulation: to reproduce, as accurately as possible, the behavior of a \emph{system under study}.
The system under study here is a physical system (either existing one or one that will be constructed).
\Cref{fig:concept_breakdown} shows the main concepts in the co-simulation domain.
To run a co-simulation, one needs a co-simulation scenario and an orchestrator algorithm.

The \emph{co-simulation scenario} points to one or more
simulation units, describing how the inputs and outputs of their models are
related.
Each \emph{simulation unit} represents a black box capable of producing behavior.
To produce behavior, the simulation unit needs to have a notion of:
\begin{compactitem}
\item a \emph{model}, which is created by the modeller based on his knowledge of the system under study;
\item a \emph{solver}, which is part of the modeling tool used by the
modeller that approximates the behavior of the model; and
\item an \emph{input
approximation}, which approximates the inputs of the model over time, to be
used by the solver;  as well as
\item \emph{input reactivity} and \emph{output reactivity}, which determine the inputs the simulation unit receives from the orchestrator. 
\end{compactitem}

The models of each simulation unit and their couplings (described by a co-simulation scenario) induce what we call an \emph{implicitly coupled model} of the system under study.
The \emph{validity} of a co-simulation is related to the validity of the implicit coupled model, that is, how well the meaning of the model represents the dynamics of interest in the system under study.
The \emph{accuracy} of a co-simulation is related to how closely the results produced by the co-simulation match the meaning of the implicitly coupled model.
The \emph{orchestrator} is responsible for producing the results of a co-simulation.
It initializes all the simulation units with the appropriate
values, sets/gets their inputs/outputs, and coordinates their progression over the simulated time.

A previous state-of-the-art survey identified two main paradigms: \emph{discrete event} (DE) and \emph{continuous time} (CT) co-simulation.
In discrete event co-simulation orchestration, the communication between simulation units is done using events (for example, in software controllers, the state tends to evolve discontinuously, as a reaction to new inputs being transmitted by the environment).
These units are characterized by their reactivity and transiency.
Reactivity means that events can change the state instantaneously (within the same simulated time), and transiency means that events can cause other events to occur instantaneously. 
In DE co-simulation, the task of the orchestrator is to forward the events published by each simulation unit to the appropriate target units and make sure the simulated time of each unit is synchronized.
In Continuous Time co-simulation, ideally, the simulation units exchange their values continuously (e.g., as in most physical systems).
However, this is not possible in practice, which as led to the development of a myriad of techniques to compensate for this limitation.

\begin{figure}[tbh]
\begin{center}
  \includegraphics[width=0.9\textwidth]{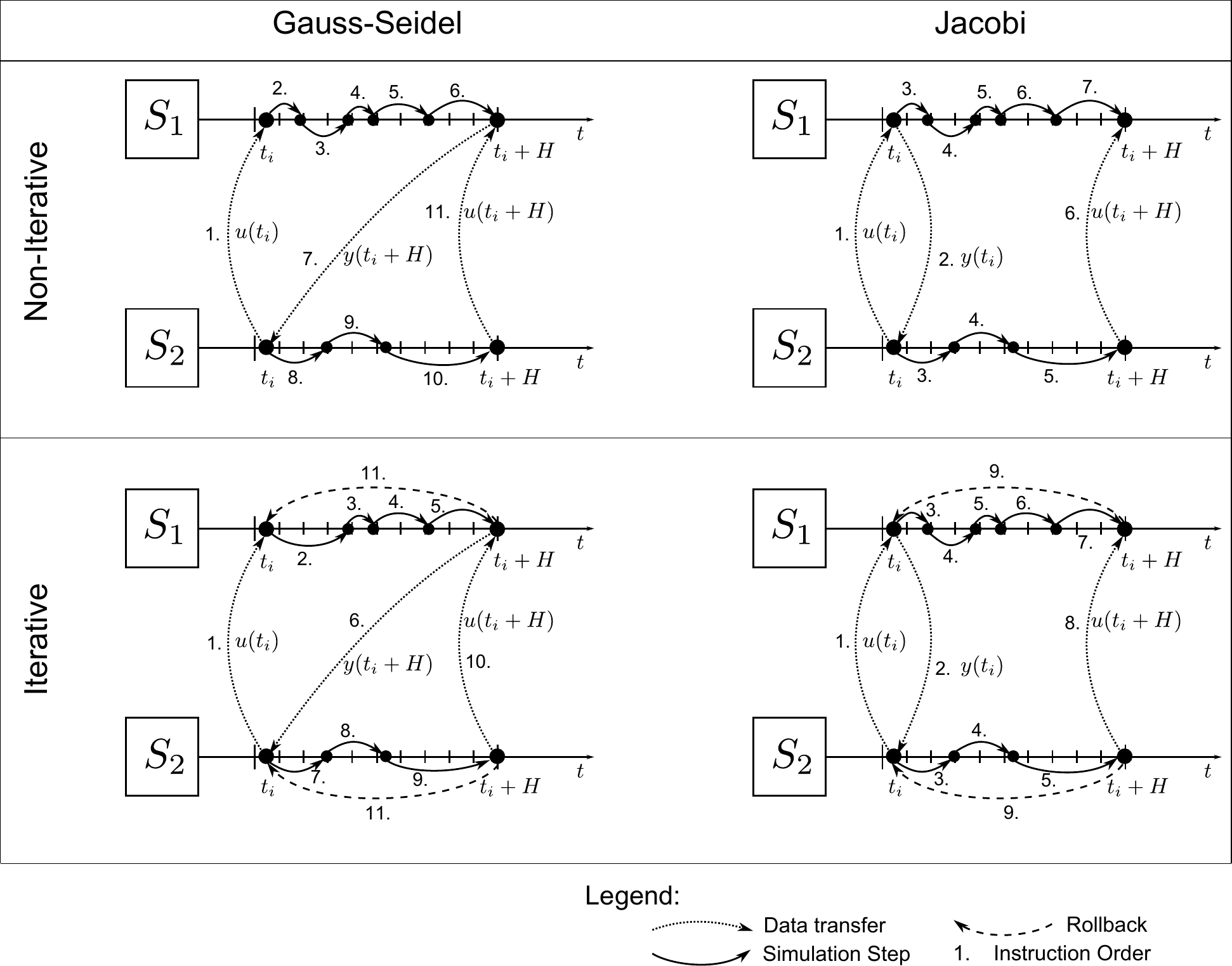}
  \caption{Overview of orchestration algorithms. Taken from \cite{Gomes2018e}.}
  \label{cosim_overview}
\end{center}
\end{figure}

In CT co-simulation, the orchestrator, after setting the appropriate inputs for the simulation units (computed from their
outputs according to the co-simulation scenario), runs a simulation for a
given interval of simulated time, by providing them with a \emph{communication step}.
The simulation units, in turn, will approximate the behavior of their
model within the interval between the current simulated time and the next
communication time, relying only on the inputs they have received at the previous communication times.
It is often the case that the simulation units will only receive more inputs at the next communication with the orchestrator,
hence, they must rely on their input approximations.
The communication step size can either be \emph{fixed} (defined before the co-simulation starts and held constant throughout its execution) or \emph{adaptive} (the orchestrator determines the best value to be used whenever they initiate computation with the simulation).
The \emph{communication approach} encodes the order in which the simulation units are given inputs and instructed to compute the next interval.
\Cref{cosim_overview} summarizes the multiple types of orchestration algorithms that use time diagrams.
In the \emph{Gauss-Seidel} approach, the orchestrator instructs one simulation unit at a time to compute until the next interval and produce outputs. Then, the orchestrator uses the most recent outputs when instructing the next unit to compute.
In the \emph{Jacobi} approach, the orchestrator instructs all units to compute the interval in parallel, setting their inputs at the end of the co-simulation step.
Finally, the orchestrator may retry the co-simulation step, using improved input estimates, which are computed from the most recent outputs.
This process can be repeated until there is no improvement in the inputs is observed (\emph{fully implicit iteration}), or a fixed number of iterations has been performed (\emph{semi-implicit iteration}). 
In the later sections, it will become clear why it is a good idea to repeat the co-simulation step.

The simplicity of DE (compared to CT) co-simulation is deceptive, as both approaches require extensive fine tuning of parameters and tolerances when they are used to approximate the behavior of continuous systems under study.
Furthermore, as \Cref{fig:concept_breakdown} shows, CT and DE co-simulation are not mutually exclusive. 
The combination of these two paradigms lead to hybrid co-simulation.

The co-simulation standard prescribes the communication interface and protocol between the simulation units and orchestrator.
Well-known standards are: Functional Mockup Interface (FMI \cite{Blochwitz2011}), High Level Architecture (HLA \cite{IEEE2010}) and DEVS~\cite{Zeigler1976}.

\section{Method and Rationale}

In this section, we describe our methodology, the expert selection process, and how the answers were handled in detail.
Furthermore, we describe how the quantitative analysis of the strengths, weaknesses, opportunities and threats (SWOT) of co-simulation was conducted.

\subsection{Delphi Method}
As a methodological foundation of this study, the Delphi method \cite{Dalkey1963} was adopted. 
The Delphi method is an empirical research method that relies on the systematic compilation of knowledge from a selected group of experts \cite{Dalkey1963,Hsu2007}.
It fosters the exploration of problems that are characterized by an incomplete state of knowledge \cite{Powell2003}, a lack historical data, or a lack of agreement within the studied field, which makes it a perfect fit to apply to co-simulation \cite{Okoli2004a}.
The aim of applying the Delphi method is to arrive at a reliable shared opinion by means of a repetitive assessment process that includes controlled feedback of opinions \cite{Landeta2006}. 
The Delphi method provides structured circumstances that ''[\ldots] can generate a closer approximation of the objective truth than would be achieved through conventional, less formal, and pooling of expert opinion''  \cite{Balasubramanian2012}. 

The Delphi study applied in the present work includes two rounds. The choice of rounds was justified by, for instance, Sommerville \cite{Somerville2008}, who argued that the changes in the participants’ views occurred in most cases during the first two rounds of the study and few insights were gained in further rounds.
The quality of the Delphi process depends on the factors of creativity, credibility, and objectivity \cite{Nowack2011}. 
To address these quality criteria we followed acknowledged guidelines that have been provided by authors such as as tose of \cite{Okoli2004a,Landeta2006,Nowack2011}.
The questions in the first round were selected based on the existing studies on co-simulation (see \Cref{sec:related_work}) and the experience of the authors of the current study.
Both rounds included qualitative (open-ended) and quantitative questions.
In the first round, the majority of questions asked were qualitative, whereas in the second round, they were quantitative. 
This ensured that the topic could be introduced in a general way in the first round. 
To see why, note that if the first round had consisted mainly of quantitative questions, there would have been an increased risk of overlooking important factors or biasing the results. 
The qualitative questions asked in the first round only addressed findings that were common among survey papers referred to above. 
In these cases, expert opinions were used to evaluate the findings of the previous surveys and enable quantitative statements and comparisons to b e made. 
The quantitative questions asked in the second round were mainly formulated based on the results of the first round and the findings reported in the recent literature (e.g., where contradictions were identified).

Regarding the number of experts, Clayton \cite{Clayton1997} indicated that fifteen to thirty experts with homogeneous expertise backgrounds or five to ten experts with heterogeneous backgrounds should be involved in a Delphi process, while Adler and Ziglio \cite{Adler1996} argued that ten to fifteen experts with homogeneous expertise backgrounds could already be considered appropriate. 

\subsection{SWOT-AHP}

The literature lacks studies that have been carried out to systematically investigate the advantages and disadvantages of co-simulation and relate them to each other. Thus, in the second round of the Delphi study, we conducted a quantitative SWOT analysis utilizing an Analytic Hierarchy Process.
A SWOT analysis is an analytic technique used to analyze the internal strengths and weaknesses, as well as the external opportunities and threats of a project, product, person, or other item \cite{Kotler2016}.
While a classical SWOT analysis may be used to pinpoint specific factors, the selected factors are not prioritized or weighted in terms of their relative importance. In practice, this complicates strategy development and means that the strategic planning process strongly depends on the individual judgments of the people involved. 
To overcome this drawback, we adopted an Analytic Hierarchy process (AHP) in our SWOT analysis. The goal of the use of this integrated SWOT-AHP method was to gain a better understanding of the relative importance of each factor. 
Therefore, experts were asked to make a pairwise comparisons and weighting of the respective factors in each category, as well as compare the categories based on a 9-point scale. 
Saaty \cite{Saaty1980} developed an AHP that is based on the eigenvalue method. 
The goal was to synthesize a pairwise comparison matrix and to get a priority for each factor in a group. 
In a first step, the relative priority of each factor in each group was calculated based on the average results of the pairwise comparisons. 
The result gained was the ''local factor priority''. 
In a second step, the ''group priority'' was calculated based on the average results of how experts assessed the priority of the individual groups. 
In a third step, the ''global factor priority'' of the respective factors was calculated by multiplying the local factor priority by the respective group priority. 
Details about the AHP method can be found in \cite{Saaty1980}.

In the first round of the Delphi study, we conducted a standard SWOT analysis. Relevant factors in the first round were selected based on an extensive literature study (see \Cref{sec:related_work}) and the experience of the authors. 
Experts were asked to select the three factors for each category that they considered the most important. 
To help validate the selected SWOT factors, an open-format question was included in each SWOT section, and experts were asked whether they considered any factors other than the ones we had selected as being more important. 
Based on these results, we selected the three most important factors per category for the second round, in which experts conducted the SWOT-AHP, performing a pair-wise comparison of all factors in the same SWOT field (i.e., stating the degree to which each factor of each pair was more important than the other). 
The experts then were asked to also compare the four SWOT groups themselves, while bearing in mind the three factors per group. 

\subsection{Expert selection and response rate}
\label{sec:expert_selection}
The Delphi method does not prescribe any particular way of selecting experts.
We used a Knowledge Resource Nomination Worksheet (KRNW) as a framework \cite{Okoli2004a}. 
The KRNW was proposed in \cite{Delbecq1975} as a general criterion that could be used to sample an expert panel by classifying the experts before selecting them in two iteration steps, to avoid overlooking any important class of experts.
This framework consists of the following five steps, detailed below:
\begin{inparaenum}[(1)]
\item \label{step:KRNW1} preparation of the KRNW;
\item \label{step:KRNW2} population of the KRNW;
\item \label{step:KRNW3} nomination of additional experts;
\item \label{step:KRNW4} ranking of experts; and 
\item \label{step:KRNW5} invitation of experts.
\end{inparaenum}

\newcommand{\KRNW}[1]{Step~(\ref{step:KRNW#1})}

In \KRNW{1}, we classified the experts according to whether they worked in \emph{academia} or \emph{industry}, as both perspectives were considered essential. 
Then, in \KRNW{2}, the \emph{academia} category was populated based on a keyword-based search in the literature on the state of the art in co-simulation (see \Cref{sec:related_work}); the \emph{industry} category was populated based on the same keyword-based search and the experience of the authors.
Afterwards, in \KRNW{3}, both categories were expanded based on the suggestions received after contacting the initial list of experts.
In \KRNW{4}, the ranking of experts was done using the number of publications in the field of co-simulation, which was obtained from Scopus.
In \KRNW{5} the final group of experts was invited to take part in the Delphi study. 
Fifteen experts were contacted for the first round; after receiving a final reminder by email, twelve completed questionnaires were returned. The response rate for the first round was, thus, 80 \%.
In the second round, we contacted seventy persons; after receiving a final reminder by email, 53 completed questionnaires were returned. The response rate for the second round was, thus, 76 \%.
We can safely state that a significant share of representatives from co-simulation experts were involved in the analysis \cite{Clayton1997,Adler1996}. 

Experts from industry who took part in the survey worked in the following sectors: energy Systems (5), software development (7), mobility (4), engineering services (1), system engineering (1), avionics, railways (1).
Experts from academia who took part in the survey work in the following fields: energy-related applications (8), software development (6),  automotive (3), computer Science (2), maritime (1),  system Engineering (1), numerical mathematics (1), system modelling and verification (1) and formal methods (1).
Some experts did not provide information about their field or sector.

\Cref{tab:rounds} summarizes the aim and approach of each round and provides the number of participants per category. 
\begin{table}[h]
\centering
\caption{Summary of method.}
\label{tab:rounds}
\begin{tabular}{lp{20em}lllll}
      &                                                                                                                                                                                            &                   & \multicolumn{4}{c}{Participants} \\ \cline{4-7} 
Round & Aim                                                                                                                                                                                        & Approach          & A     & I     & ND    & Total    \\ \hline
1     & Identification of research needs, SWOT factors, limitations and possible extension.                                                              & Qualitative       & 7     & 2     & 3     & 12       \\
2     & Evaluation of the result from the first round and development of in-depth discussions on the key aspects. Test on convergence the identified factors, themes and scenarios  & \makecell{Semi- \\ quantitative}. & 24    & 19    & 10    &    53      \\
\end{tabular}%
\end{table}

\subsection{Presentation of the results}
A content analysis was performed following the method of Mayring to analyze the qualitative answers \cite{Mayring2004}.
Authors of scientific literature have conducted controversial discussions about which statistical measures are suitable for the interpretation of results of a survey, such as Likert-scales. 
Hallowell and  Gambatese \cite{Hallowell2010} argued that results should be reported in terms of the median rather than the mean, because the median response is less likely to be affected by biased responses. 
The median is the middle observation in a sorted list of data, separating the upper half from the lower half of a dataset.
Sachs \cite{Sachs1997} argued that the interpolated median is more precise than the normal median, because it is better to consider the frequencies of answers within one category in comparison to all answers. 
The interpolated median is used to adjusts the median upward or downward within the lower and upper bounds of the Median ($M$), in the direction in which the data are more heavily weighted. The interpolated median ($IM$) is calculated as follows:

\begin{equation}\label{eq:first}
IM =
\begin{cases} 
      M & \text{if $n_2=0$}, \\
      M-0.5+ \frac{0.5 \cdot N-n_1}{n_2} & \text{if $n_2\neq 0$}
   \end{cases}
\end{equation}

where $N$ is the total number of responses to the question, $n_1$ iy the number of scores strictly less than $M$ and $n_2$ is the number of scores equal to $M$.
In order to provide a transparent presentation of the results, (i) all results are displayed in detail in a bar chart in the appendix and (ii) in Section 3, all results are discussed using mean, median and interpolated median values.
Remarkable agreement or differences among experts (or groups of them) are highlighted when identified.

\subsection{Threats to validity and limitations of the study}

Detailed discussion about the threats to validity in Delphi studies can be found in \cite{Hasson2000}.
The selection of experts from academia was done based strictly on the number of publications listed in Scopus.
There is an ongoing discussion about how to compare the scientific impact among researchers. 
While some indices are well-suited for comparing researchers within the same field, this is not the case for comparing different fields. 
Since co-simulation is an interdisciplinary field of research, the selection of experts in this work can be seen as a threat to validity. 
The ranking of experts from industry was done based on the number of publications listed in Scopus. In addition, we selected experts from industry who we knew have been working with co-simulation for a long time and who have theoretical and practical knowledge in the field of co-simulation. It can be regarded as a limitation regarding the representativity of the results; however, the responses of the experts indicated that they indeed were well experienced
This selection process ensured that also experts from industry whose focus is not on scientific publishing participated in the study.


\section{Results and Discussion}


In this section, we present the key findings from the Delphi study and the SWOT-AHP analysis.
In the results below, most questions are multiple choice, and the options available were collected during the first round of the Delphi study. 
To accommodate for additional answers, an extra open field was provided.
These open answers, where applicable, are displayed under the \emph{Other} category.

\subsection{Simulator and Co-simulation Characterization}

\begin{figure}[h!]
\centering
\includegraphics[width=1\textwidth]{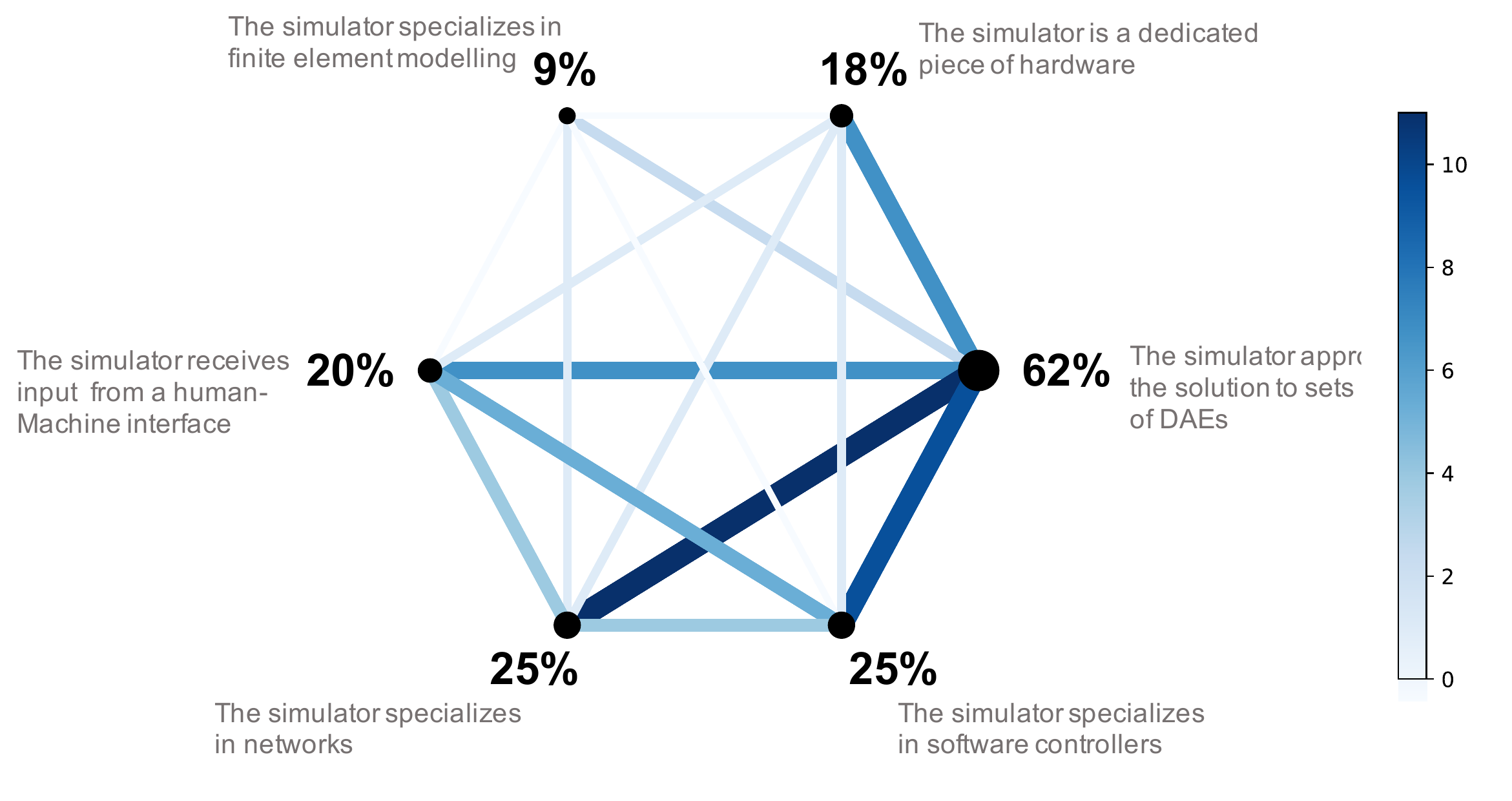}
\caption{Answers to the question: ``which properties apply to the simulators \ldots?''. 
Each node represents a property. 
The size of each node is proportional to the number of positive responses to the corresponding property. 
Moreover, the thickness of the edge-connections nodes $x$ to nodes $y$ indicates that the same expert gave positive reply to both property $x$ and $y$.
Note that the latter does not imply (and neither neglect) the different properties to apply in one and the same co-simulation.
}
\label{fig:CoSim}
\end{figure}

In order to analyze the purpose for which experts used co-simulation, experts were asked to select the properties that apply to the simulators with which they have worked in co-simulation.
As can be seen in  \Cref{fig:CoSim}, the majority of the simulators being used in co-simulation represented sets of differential equations. 
Still, between $18\%$ and $25\%$ of the experts used simulators as ``specialized in networks'', as ``specialized in software controllers'', 
as ``a dedicated piece of hardware'' or as ``receiving input from a human machine interface''.
 
The properties that were not pre-defined in the questionnaire represented a minority of answers: one expert used co-simulation to prove a theorem and another, to solve partial differential equations using finite volume methods.
These results indicate that the first round of the Delphi study was successful in that the uses of the simulators could be characterized, and determine that co-simulation was used for many different applications.

\subsection{Dissemination channels}
\label{sec:Diss}

\begin{figure}[h!]
\centering
\includegraphics[width=1\textwidth]{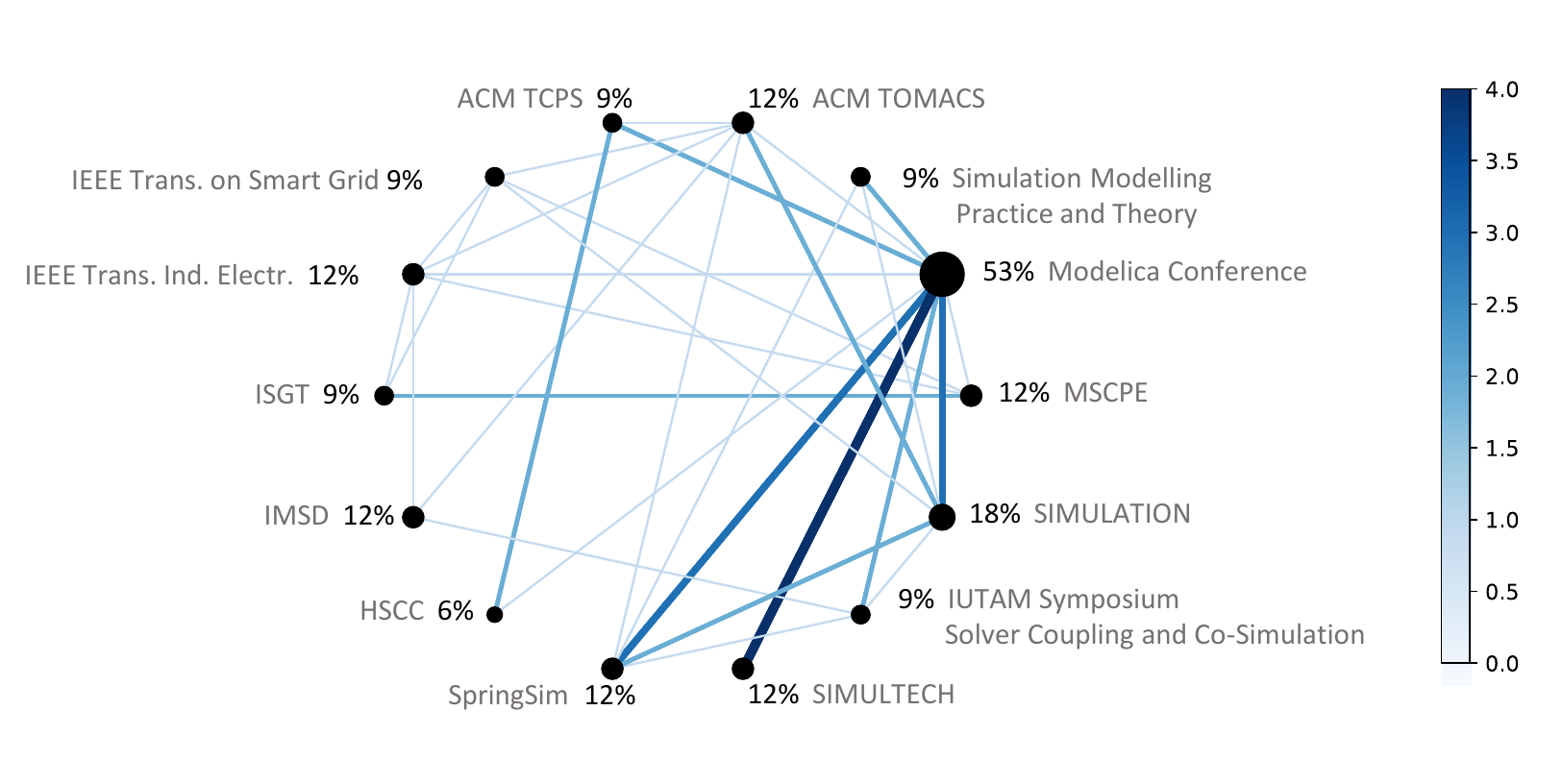}
\caption{
Experts were asked to mark the three most important scientific sources they used to disseminate their work. The numbers next to the nodes correspond to the ... \% of positive responses; the size of the nodes is also proportional to number of positive responses. The statements upon which the respective experts agreed were connected. MSCPE = Workshop on Modeling and Simulation of Cyber-Physical Energy Systems; ISGT = IEEE Conference on Innovative Smart Grid Technologies; IMSD = International Conference on Multibody System Dynamics; HSCC
= Conference on Hybrid Systems: Computation and Control. Conferences mentioned only once by experts are not shown; these included the Conference of the IEEE Industrial Electronics Society, IEEE transactions on power delivery, IEEE Power \& Energy Society General Meeting, International Association of Applied Mathematics and Mechanics, Problems in Science and Engineering, European Community on Computational Methods in Applied Sciences and Workshop on Co-simulation of Cyber Physical Systems.
}
\label{fig:Diss}
\end{figure}

To identify the main dissemination channels, experts were asked to name the three most important scientific sources used to disseminate their work. The results are shown in \Cref{fig:Diss}. 
The Modelica Conference was cited as by far the most important channel for experts used to disseminate their work.
The FMI has been one of two key topics in this conference, suggesting that this result 
is co-related with the fact that the FMI is considered to be the most promising standard for co-simulation (see \Cref{sec:StandTools}).
The dissemination channels suggested by the experts are highly heterogeneous, which underlines the assumption that co-simulation is indeed a multi-disciplinary research field. 

\subsection{Established Standards and Tools}
\label{sec:StandTools}

\begin{figure}[h!]
\centering
\includegraphics[width=1\textwidth]{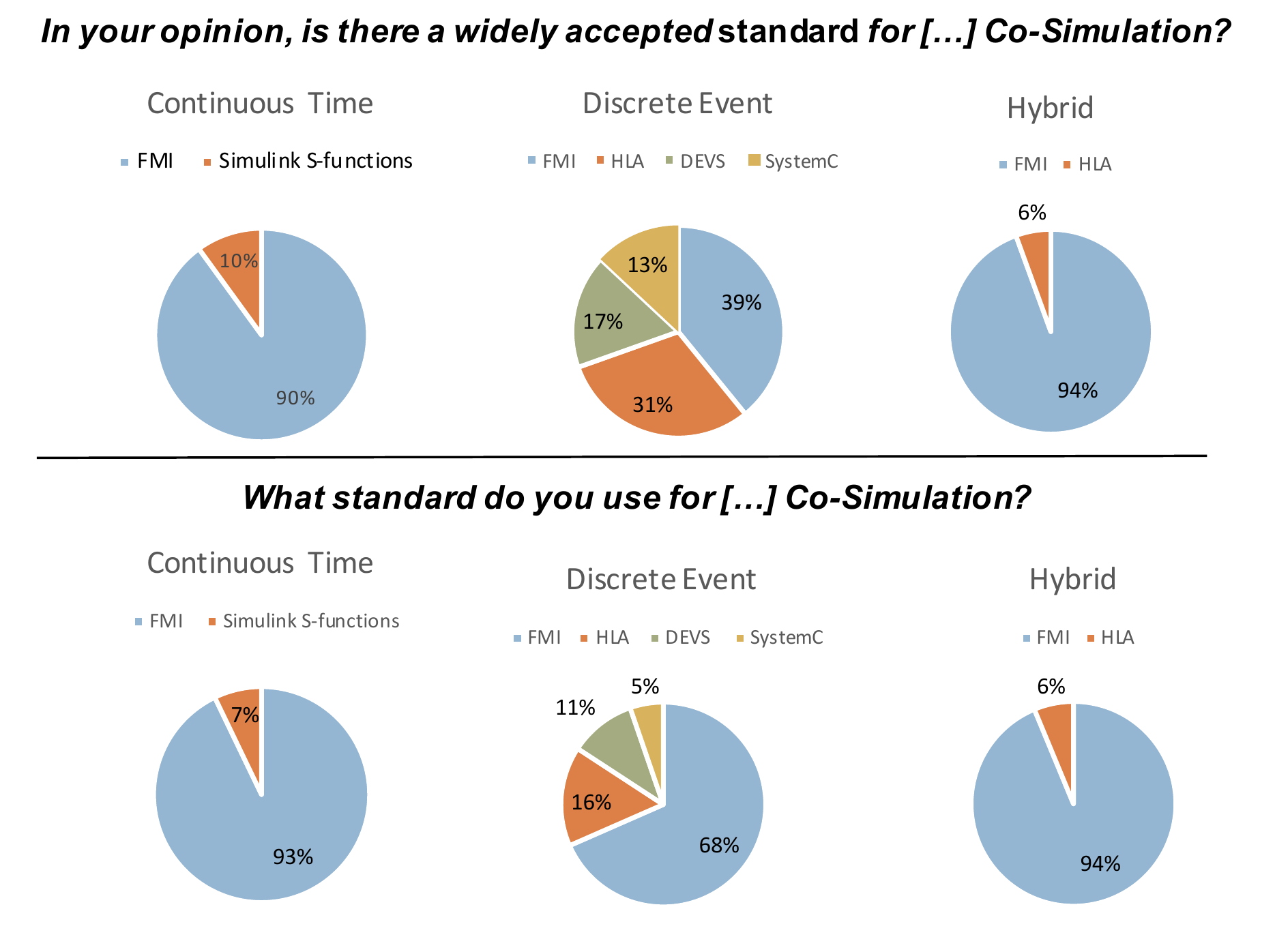}
\caption{Widely accepted and used standards for co-simulation. Depending on the sub figure, the brackets [...] correspond to ``Continuous Time'', ``Discrete Event'' or ``Hybrid''  }
\label{fig:Standards}
\end{figure}

To identify promising standards for continuous time, discrete event and hybrid co-simulation, we asked experts (i) to give their opinion on widely accepted standards and describe (ii) what standard they used for co-simulation. 
We would like to point out that no generally valid statements can be derived here as to which standards are widely used in industry and academia; the sample size is too small for this and the influence of possible biases on the selection of experts is too high.  
The results are summarized in \Cref{fig:Standards}.

As can be seen in the figure, the FMI standard is by far the most commonly used standard for any kind of co-simulation. 

While the responses for "widely accepted standards" and "standards which experts use" were similar for continuous time and hybrid co-simulation, a different picture emerged for discrete event co-simulation. FMI was described as widely accepted for discrete event co-simulation by 39 \% of the experts, however, 68 \% of the experts used FMI for discrete event co-simulation. 
A dedicated empirical study, similar to the one presented here, was performed to identify challenges/barriers to the adoption of the FMI standard \cite{Schweiger2018b}.
The main results of that study are summarized in Table \ref{tab:FMI}.

\begin{table*}[t]
\caption{Expert assessment of current barriers for FMI. Based on a Seven-point Likert scale. Modified from \cite{Schweiger2018b}.
}
\label{tab:FMI}
\centering
\includegraphics[width=0.99 \textwidth]{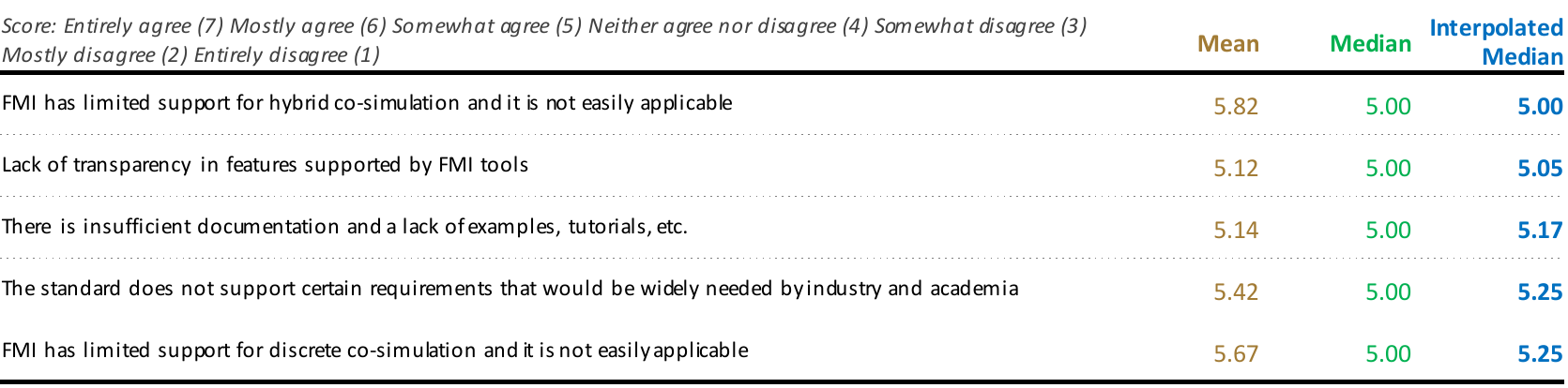}
\end{table*}

\begin{figure}[h!]
\centering
\includegraphics[width=0.7\textwidth]{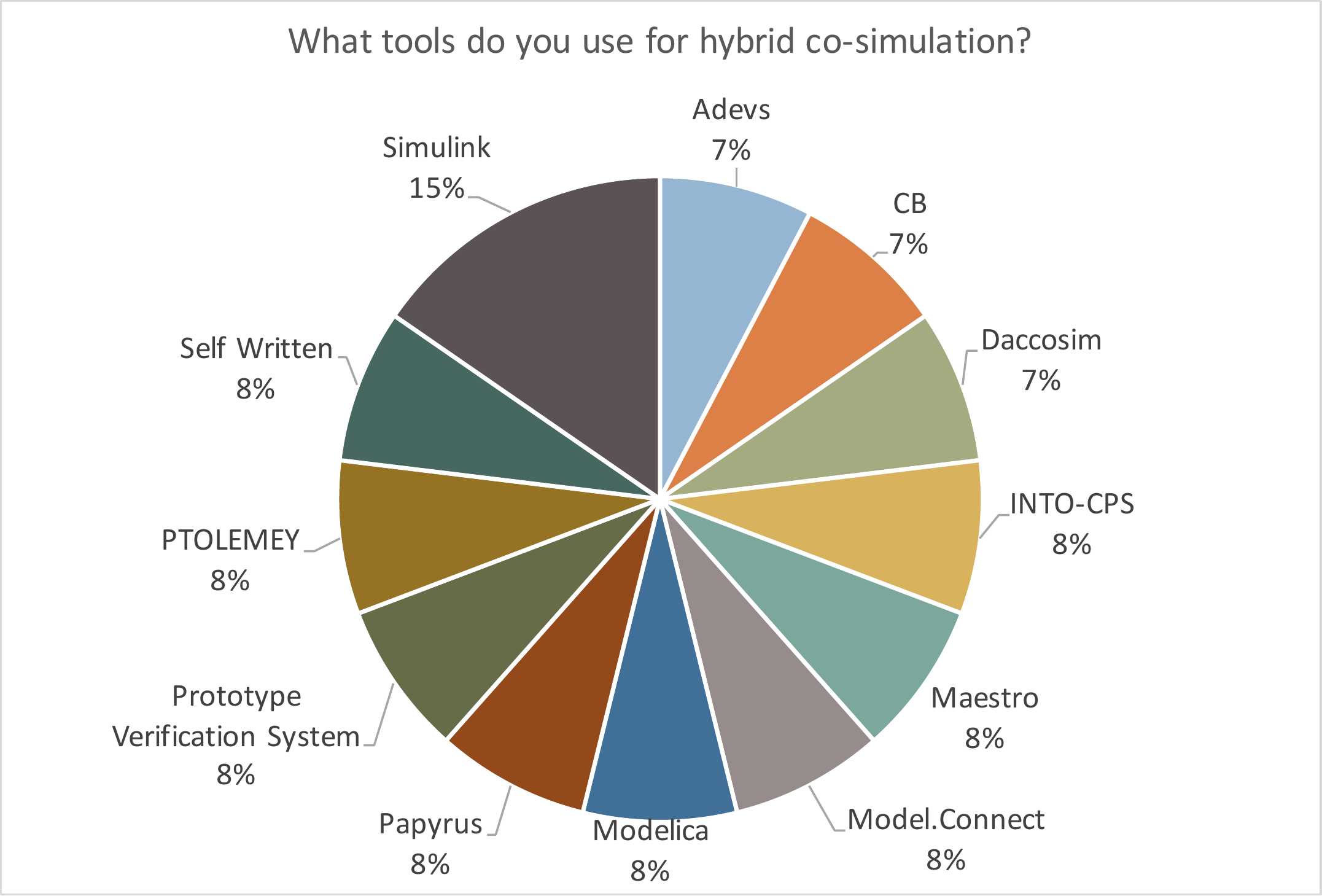}
\caption{Tools that experts use for hybrid co-simulation. }
\label{fig:ToolsH}
\end{figure}

In addition to promising standards, experts were asked which tools they used for co-simulation. 
The most common tools used for continuous time co-simulation were Modelica tools and Matlab/Simulink. 
The use of Modelica toos was cited by  40 \% of the experts and about 24 \% of the experts mentioned that they used Matlab/Simulink. 
For discrete event and hybrid co-simulation, no tool was significantly more frequently mentioned than others.
The detailed results can be found in the Appendix (Tables \ref{fig:ToolsC}, \ref{fig:ToolsD} and \ref{fig:ToolsH}).
While only seven different tools were listed for CT co-simulation, thirteen were listed for DE twelve 12 for hybrid co-simulation.

\subsection{Current challenges}
\label{sec:CurrChall}

\begin{figure}[h!]
\centering
\includegraphics[width=0.8\textwidth]{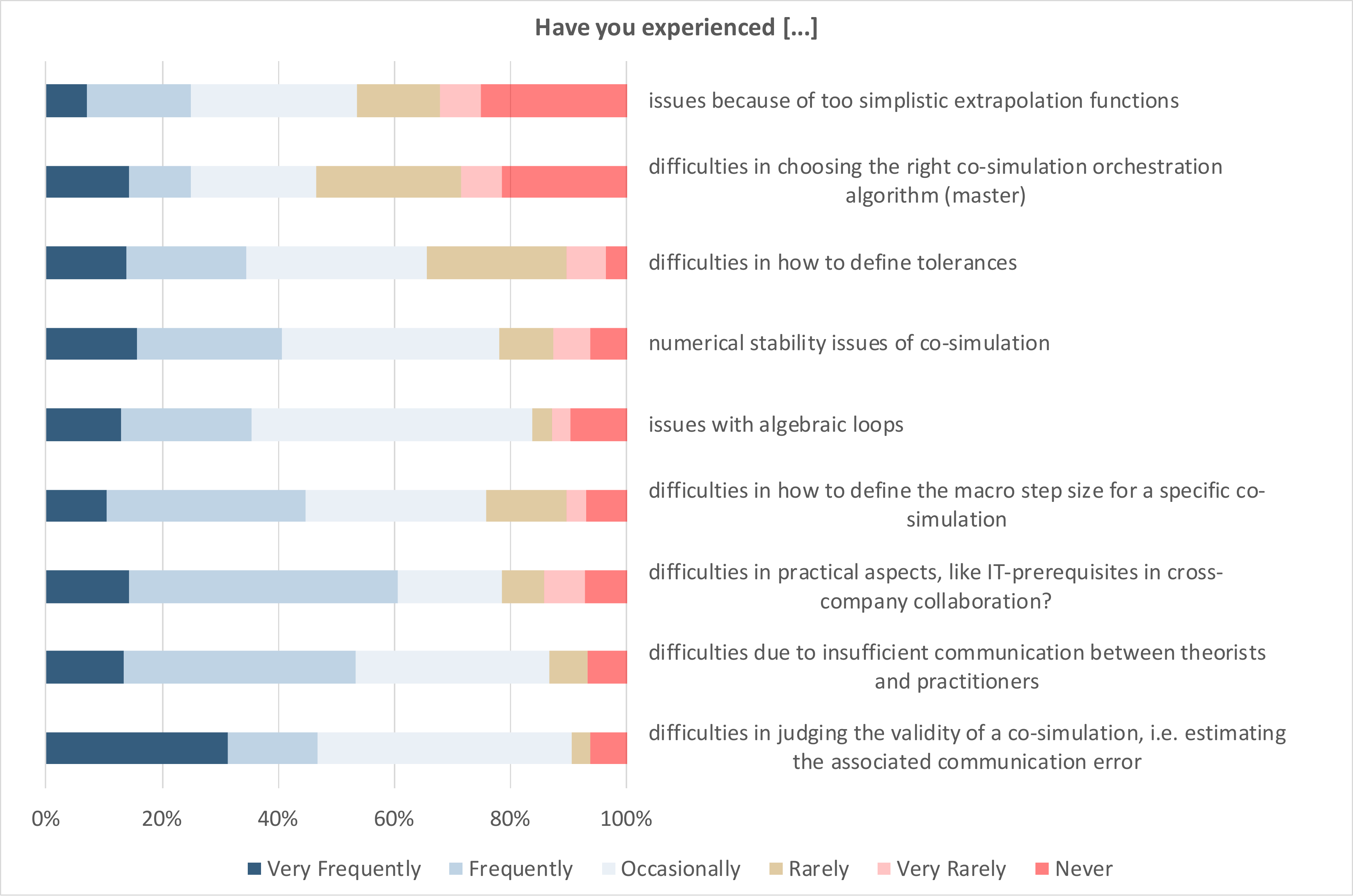}
\caption{Current challenges}
\label{fig:challenges}
\end{figure}

In the first round of the Delphi study, experts commented on current challenges. 
Based on these responses and the state-of-the-art surveys, we formulated several statements regarding personal experiences.
In the second round of the Delphi Study, we posed these statements as questions (e.g., ``Have you experienced\ldots'').
The experts then used a 6-point Likert scale that ranged from from $1 = \text{``very frequently''}$ to $6 = \text{``never''}$. 
\Cref{fig:challenges} shows the response count, and \Cref{tab:LS} summarizes the responses sorted according to how often experts experience each challenge. 
A detailed discussion of the individual challenges goes beyond the scope of this survey. 
However, appropriate references are provided next to each challenge.
\begin{table}[htbp]
  \centering
  \caption{Experts' assessments: Current challenges. Score: Very Frequently (6) Frequently (5) Occasionally (4) Rarely (3) Very Rarely (2) Never (1).}
  \label{tab:LS}%
    \begin{tabularx}{1\linewidth}{
        X
        >{\color{black}}r
        >{\color{black}}r
        >{\color{red}}r
        <{\color{black}}
    }
    & \thead{Mean} & \thead{Median} & \thead{Interp.\\Median} \\
    \midrule
        Difficulties in practical aspects, like IT-prerequisites in cross-company collaboration.  & 4.7   & 5.0   & 4.7 \\
            Difficulties due to insufficient communication between theorists and practitioners.   & 4.4   & 5.0   & 4.6 \\
    Difficulties in judging the validity of a co-simulation.   & 4.6   & 4.0   & 4.4 \\
    Difficulties in how to define the macro step size for a specific co-simulation \cite{Benedikt2013b,Busch2011,Gomes2017}.  & 4.3   & 4.0   & 4.3 \\
        Numerical stability issues of co-simulation \cite{Busch2016,Gomes2018,Arnold2010}.  & 4.4   & 4.0   & 4.3 \\
    Issues with algebraic loops \cite{Kubler2000,Gomes2017}. & 4.2   & 4.0   & 4.2 \\

    Difficulties in how to define tolerances.  & 4.3   & 4.0   & 4.0 \\
    Issues because of too simplistic extrapolation functions.  & 3.5   & 4.0   & 3.6 \\
    Difficulties in choosing the right co-simulation orchestration algorithm (master).  & 3.6   & 3.0   & 3.4 \\
    \bottomrule
    \end{tabularx}%
\end{table}%

The results indicate that practical aspects (as opposed to scientific problems) dominate the problems encountered when conducting co-simulation.
The difficulties encountered when judging the validity of a co-simulation present pertinent challenges, already important in the simulation field \cite{Spiegel2005,Denil2017} and aggravated by the black-box nature of co-simulation. 
Most challenges (all except simplistic extrapolation functions and difficulties in choosing the correct orchestration algorithm) were assessed by the experts with an interpolated median value greater or equal to four, implying at least occasional occurrence. 
The experts, thus, confirm the challenges identified in the first round and from the state of the art. 
Many experts identified having difficulty choosing the right macro step size, defining tolerances and with numerical stability. From these responses, we conclude that there is a need for frameworks that provide suitable suggestions to ease the choices for the user. 
In particular, we assume that many user have significantly less know-how in the areas in practice than the experts interviewed in this work.

\subsection{Research needs}
\label{sec:ResNeed}

\begin{figure}[h!]
\centering
\includegraphics[width=0.8\textwidth]{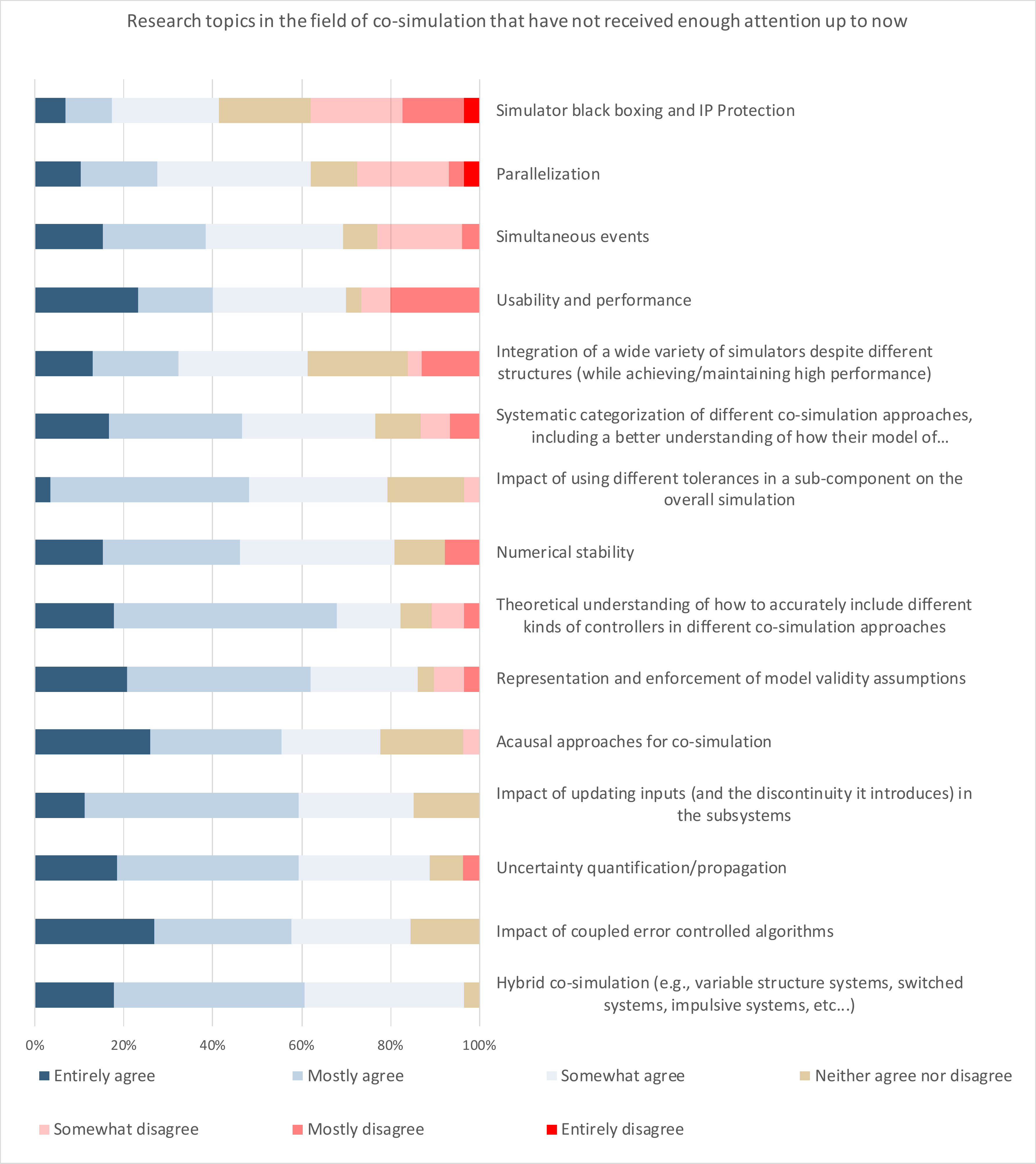}
\caption{Research needs. }
\label{fig:ResearchNeeds}
\end{figure}

Experts were asked about research topics in the field of co-simulation that have not received enough attention up until now.
\Cref{fig:ResearchNeeds} summarizes the response count on a 7-point scale from ``Entirely Disagree'' to ``Entirely agree''.
Furthermore, \Cref{tab:Research} shows the same data, sorted in ascending order of topics that \emph{have not} received enough attention until now. 


\begin{table}[htbp]
  \footnotesize
  \centering
  \caption{Experts assessments: Research needs. Score: Entirely agree (7) Mostly agree (6) Somewhat agree (5) Neither agree nor disagree (4) Somewhat disagree (3) Mostly disagree (2) Entirely disagree (1).}
  \label{tab:Research}%
    \begin{tabularx}{1\linewidth}{
        X
        >{\color{black}}r
        >{\color{black}}r
        >{\color{red}}r
        <{\color{black}}
    }
    & \thead{Mean} & \thead{Median} & \thead{Interp.\\Median} \\
    \midrule
    Theoretical understanding of how to accurately include different kinds of controllers in different co-simulation approaches  & 5.5   & 6.0   & 5.9 \\
    Representation and enforcement of model validity assumptions \cite{Spiegel2005,Denil2017}  & 5.6   & 6.0   & 5.8 \\
    Hybrid co-simulation (e.g., variable structure systems, switched systems, impulsive systems, etc...) \cite{Cremona2017a,Gomes2018}  & 5.8   & 6.0   & 5.8 \\
    Impact of coupled error controlled algorithms \cite{Busch2011,Hafner2013} & 5.7   & 6.0   & 5.8 \\
    Uncertainty quantification/propagation \cite{Bouissou2013,Lawrence2016} & 5.6   & 6.0   & 5.8 \\
    Impact of updating inputs (and the discontinuity it introduces) in the subsystems \cite{Busch2016,Gomes2017c}.  & 5.6   & 6.0   & 5.7 \\
    Acausal approaches for co-simulation \cite{Sadjina2017}  & 5.6   & 6.0   & 5.7 \\
    Impact of using different tolerances in a sub-component on the overall simulation \cite{Arnold2014}  & 5.3   & 6.0   & 5.5 \\
    Numerical stability \cite{Busch2010,Gomes2017d,Gomes2018c} & 5.3   & 5.0   & 5.4 \\
    Systematic categorization of different co-simulation approaches, including a better understanding of how their model of computations and requirements overlap and differ \cite{Thule2018} & 5.2   & 5.0   & 5.4 \\
    Usability and performance  & 4.9   & 5.0   & 5.2 \\
    Simultaneous events \cite{Broman2018}  & 5.0   & 5.0   & 5.1 \\
    Integration of a wide variety of simulators despite different structures (while achieving/maintaining high performance) \cite{Gomes2018a}  & 4.8   & 5.0   & 4.9 \\
    Parallelization \cite{Saidi2016,Thule2016}  & 4.6   & 5.0   & 4.9 \\
    Simulator black boxing and IP Protection \cite{Blochwitz2011} & 4.1   & 4.0   & 4.1 \\
    \bottomrule
    \end{tabularx}%
   
\end{table}%
Most research needs (all except simulator black boxing and IP protection) are assessed by the experts with a interpolated median value greater 4.5, corresponding to at least "Somewhat agree". Seven research needs were rated with an interpolated median score of greater or equal to 5.5 which corresponds to at least "Mostly agree".
The experts thus confirm the research needs identified in the first round and from the existing surveys.
In the context of hybrid co-simulation, an expert mentioned that there is only limited awareness about the problems that can arise in hybrid co-simulation; in many cases, it is difficult for user to understand whether problems arise due to shortcomings in standards, tool implementation, or usage. 
Another expert stressed that the fundamental question of what hybrid co-simulation is and what it should be able to do is a controversial one.
Is the intention to allow the same flexibility with hybrid co-simulation as there is in monolithic simulation (with everything that this entails) ir us the intention to couple large subsystems? This expert concluded that the two different views have extremeley different requirements with respect to hybrid co-simulation, and this is a largely unexplored topic that needs more research regarding numerical properties.




\subsubsection{Miscellaneous}
Experts were asked the extend to which they agreed on several statements. 
The 7-point Likert scale was used to measure the responses (Entirely agree =7 to Entirely disagree = 1). 
Experts mostly agreed with the statement \textit{"For academia it is difficult to experiment with different co-simulation approaches as there is a huge learning curve: in terms of learning the specification and also gaining access to models as well as being able to make changes to existing approaches and test new ideas"} ($Mean = 5.5, M = 6.0, IM = 5.8)$) 
and somewhat agreed with the statements \textit{"A clearer categorization of different co-simulation approaches would help for your particular field of work"} ($Mean = 5.8, M = 5.0, IM = 5.1)$) 
and \textit{"The major benefit of co-simulation is to increase performance, when compared to a monolithic simulation"} ($Mean = 4.7, M = 5.0, IM = 4.7)$); they neither agreed nor disagree with the statement \textit{"A acausal approaches can boost the use of co-simulation in your field"} ($Mean = 5.1, M = 4.0, IM = 4.3)$).




\subsection{SWOT-AHP}
\label{sec:SWOT}

\begin{figure}[h!]
\centering
\includegraphics[width=1\textwidth]{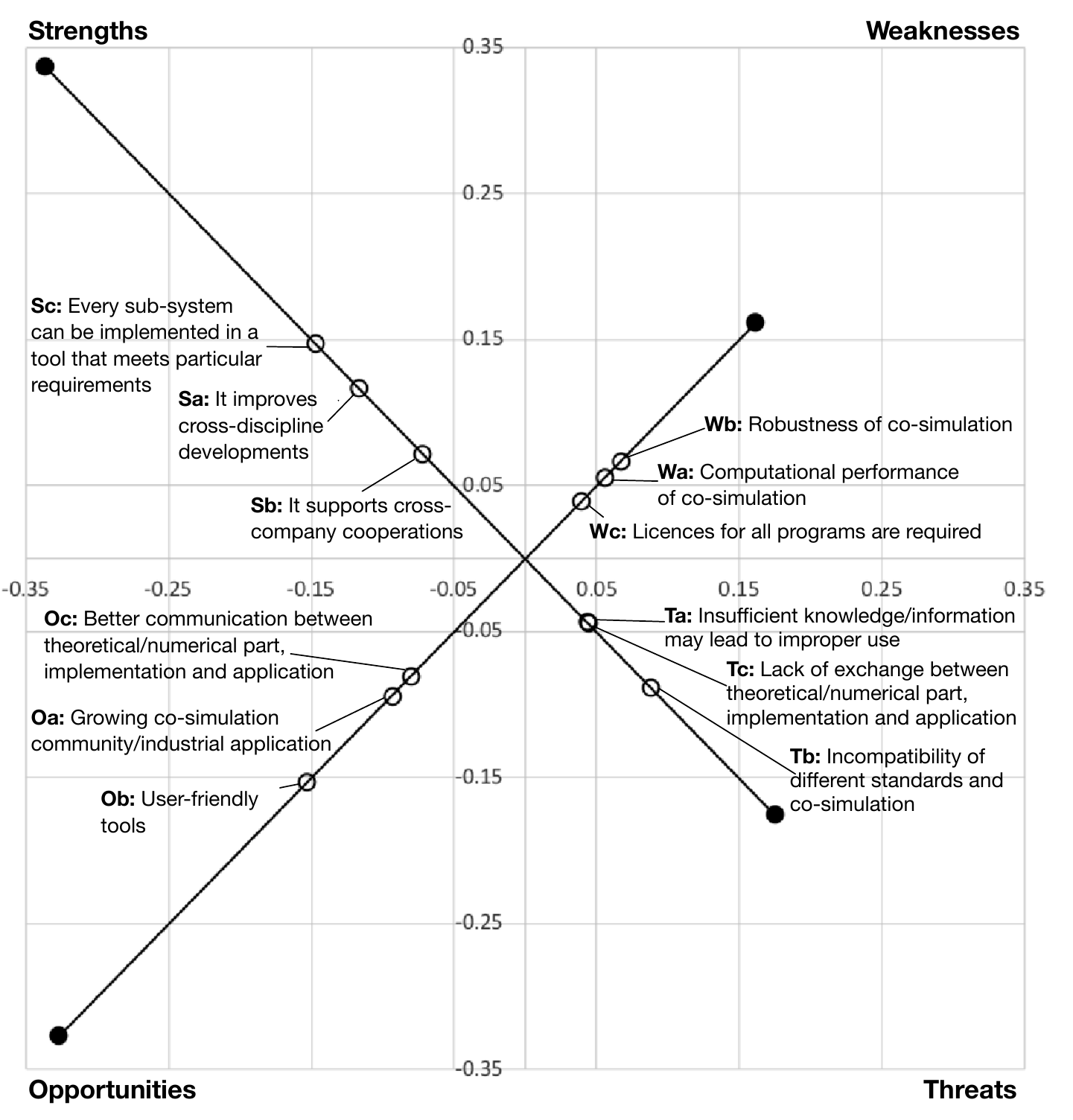}
\caption{SWOT-AHP for the research needs. 
Black dots indicate group priorities, circles indicate global factor priorities.}
\label{fig:SWOT}
\end{figure}

The results of the SWOT-AHP analysis are presented in \Cref{tab:swot_ahp} and in \Cref{fig:SWOT}. 
The factors for each group are given along the lines in the four sectors. 
The lengths of the lines indicate the group priority and, respectively the relative overall importance of the four SWOT-groups. 
The three circles per group indicate the global factor priorities; the longer the distances between the respective group/factor and the origin, the higher the overall importance assigned this group/factor.

\begin{table}[htbp]
\caption{Result SWOT-AHP}
\label{tab:swot_ahp}
\centering
\scriptsize
\begin{tabular}{p{0.4\textwidth}p{0.1\textwidth}p{0.1\textwidth}p{0.1\textwidth}p{0.1\textwidth}}
\hline
\textbf{SWOT Factors} & \textbf{CR} & \textbf{group priority} & \textbf{local priority (rank)} & \textbf{global priority (rank)} \\ \hline
\textbf{Strengths (internal)} & 0.085 & 0.34 &  &  \\
\textit{Sa: It supports cross-discipline developments}  &  &  & 0.35 (2) & 0.117 (3) \\
\textit{Sb: It supports cross-company cooperations}   &  &  & 0.21 (3) & 0.072 (7) \\
\textit{Sc: Every sub-system can be implemented in a tool that meets the particular requirements for the domain, the structure of the model and the simulation algorithm}    & &  & 0.44 (1) & 0.148 (2) \\ \hline
\textbf{Weaknesses (internal)}   & 0.013 & 0.16 &  &  \\ \hline
\textit{Wa: Computational performance of co-simulation compared to monolithic simulation}   &  &  & 0.34 (2) & 0.056 (9) \\
\textit{Wb: Robustness of co-simulation compared to monolithic simulation}   &  &  & 0.41 (1) & 0.067 (8) \\
\textit{Wc: Licenses for all programs are required to couple different simulation programs}   &  &  & 0.24 (3) & 0.039 (12) \\ \hline
\textbf{Opportunities (external)}   & 0.003 & 0.33 &  &  \\ \hline
\textit{Oa: Growing co-simulation community/growing industrial adoption}   &  &  & 0.29 (2) & 0.094 (4) \\
\textit{Ob: User-friendly tools (pre-defined master algorithms, integrated error estimation, sophisticated analysis to determine best parametrization of solvers and master algorithms)}   &  &  & 0.47 (1) & 0.153 (1) \\
\textit{Oc: Better communication between theoretical/numerical part, implementation and application/industry}   &  &  & 0.25 (3) & 0.080 (6) \\ \hline
\textbf{Threats (external)}   & 0.003 & 0.18 &  &  \\ \hline
\textit{Ta: Insufficient knowledge/information of user in co-simulation may lead to improper use}   &  &  & 0.28 (3) & 0.088 (5) \\
\textit{Tb: Incompatibility of different standards and co-simulation approaches}  &   &  & 0.41 (1) & 0.043 (11) \\
\textit{Tc: Lack of exchange/cooperation between theoretical/numerical part, implementation and application/industry.}   &  &  & 0.31 (2) & 0.044 (10) \\ \hline
\end{tabular}%
\end{table}

The results of the SWOT-AHP analysis indicate that factors for strengths and opportunities predominate. The four factors with the highest global priorities fell within the Strengths and Opportunities group. The factor with the highest global priority is the external opportunity of ``user-friendly tools including pre-defined master algorithms, integrated error estimation, etc.''. 
The factor with the second-highest global priority is the internal strength that ``sub-systems can be implemented in a tool that meets the particular requirements for the domain, the structure of the model and the simulation algorithm''. 
The factor with the third-highest global priority is the internal strength 
that ``co-simulation supports cross-discipline developments''. 
Some experts mentioned additional SWOT factors. 
As a strength, some experts mentioned that ``parallel modeling and simulation can reduce the overall modeling and simulation time''. 
Another strength identified was ``co-simulation approaches supports modularity and the reuse of components.'' 
An expert stated the ``lack of sufficiently strong theory'' as a weakness of co-simulation. 
In the group opportunities, an expert mentioned the ``integration of tools for the application of formal methods.'' 
One expert pointed out that a threat could be that ``some big companies may be actively against the widespread use of co-simulation''.

An interesting outcome is that the groups of Strengths and Opportunities were reviewed as much more important than the Weaknesses and Threats.
This was due to the priority assigned to the first two groups, which was approximately twice as high as the latter two. 
The consistencies of the pairwise comparisons was checked. All consistency ratios are below $0.1$. It can be concluded, that the results are consistent.

\section{Conclusions and outlook}

The present paper presents an expert assessment on co-simulation, adressing the social and empirical aspects and placing a focus on promising standards and tools, current challenges and research needs. 
As a methodological foundation of this study, the Delphi method was adopted.
Furthermore, a quantitative analysis of the SWOT of co-simulation utilizing the Analytic Hierarchy Process was conducted.
The authors consider the following findings from the empirical data as the most important:

\begin{compactitem}
\item Experts consider the FMI standard as the most promising standards for continuous time, discrete event  and  hybrid  co-simulation;
\item Experts frequently have difficulties dealing with practical aspects, like IT-prerequisites in cross-company collaboration, and encounter problems due to
insufficient communication between theorists and practitioners.
\item The most important research needs identified by experts are: (i) theoretical understanding of how to accurately include different kinds of controllers in different co-simulation approaches, (ii) validity aspects, (iii) hybrid co-simulation (iv) accuracy aspects and (v) acausal approaches;  
\item The highest ranked difficulty relates to practical aspects while the highest ranked research need related to theoretical understanding. This is not a contradiction; this insight may help for making co-simulation wide-spread;
\item The  results  of  the  SWOT-AHP  analysis  indicate that factors for strengths and opportunities predominate.
The experts assign the highest important to the need for user-friendly tools including pre-defined master algorithms, integrated error estimation, etc.
\end{compactitem} \leavevmode
\\
Statistical tests were conducted to determine differences in the perceptions of experts from industry and academia regarding the current challenges and open research topics; no significant difference were observed.
We refrained from testing more complex hypotheses in this study, due to the number of answers and the non-probability sampling approach taken.
However, the results of this study can be used as a basis for a follow-up, purely deductive study in which various hypothesis can be tested with a larger sample

It is our hope that the results of this study will increase transparency and facilitate the structured development of co-simulation standards and tools.






\section*{Acknowledgement}

We sincerely thank all experts who participated in the study.
The research was supported by ECSEL JU under the project H2020 737469 AutoDrive - Advancing fail-aware, fail-safe, and fail-operational electronic components, systems, and architectures for fully automated driving to make future mobility safer, affordable, and end-user acceptable. AutoDrive is funded by the Austrian Federal Ministry of Transport, Innovation and Technology (BMVIT) under the program "ICT of the Future" between May 2017 and April 2020. More information \url{https://iktderzukunft.at/en/} \includegraphics[width=1.1cm]{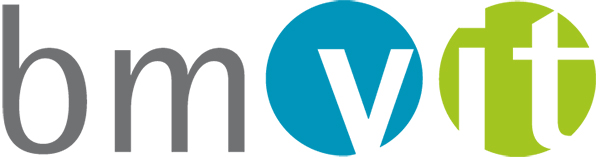}. The  financial  support by  the  Austrian  Federal  Ministry  for Digital  and  Economic Affairs and the National Foundation for Research, Technology and Development is gratefully acknowledged.

\section{References}

\bibliographystyle{plain}
\bibliography{Lit10}


\section{Appendix}
\label{sec:Appendix}

\begin{figure}[h!]
\centering
\includegraphics[width=0.7\textwidth]{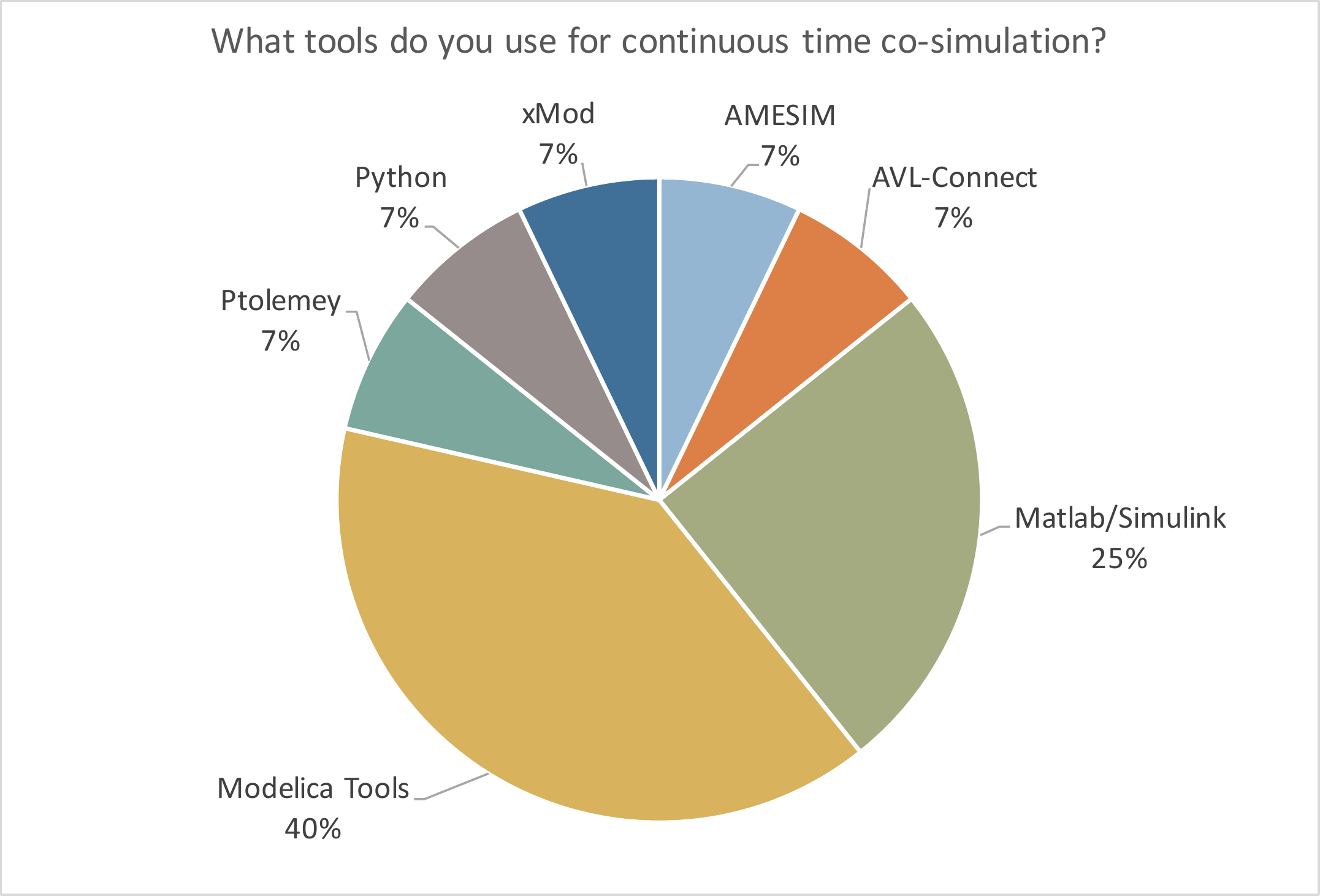}
\caption{Tools that experts use for continuous time co-simulation}
\label{fig:ToolsC}
\end{figure}

\begin{figure}[h!]
\centering
\includegraphics[width=0.7\textwidth]{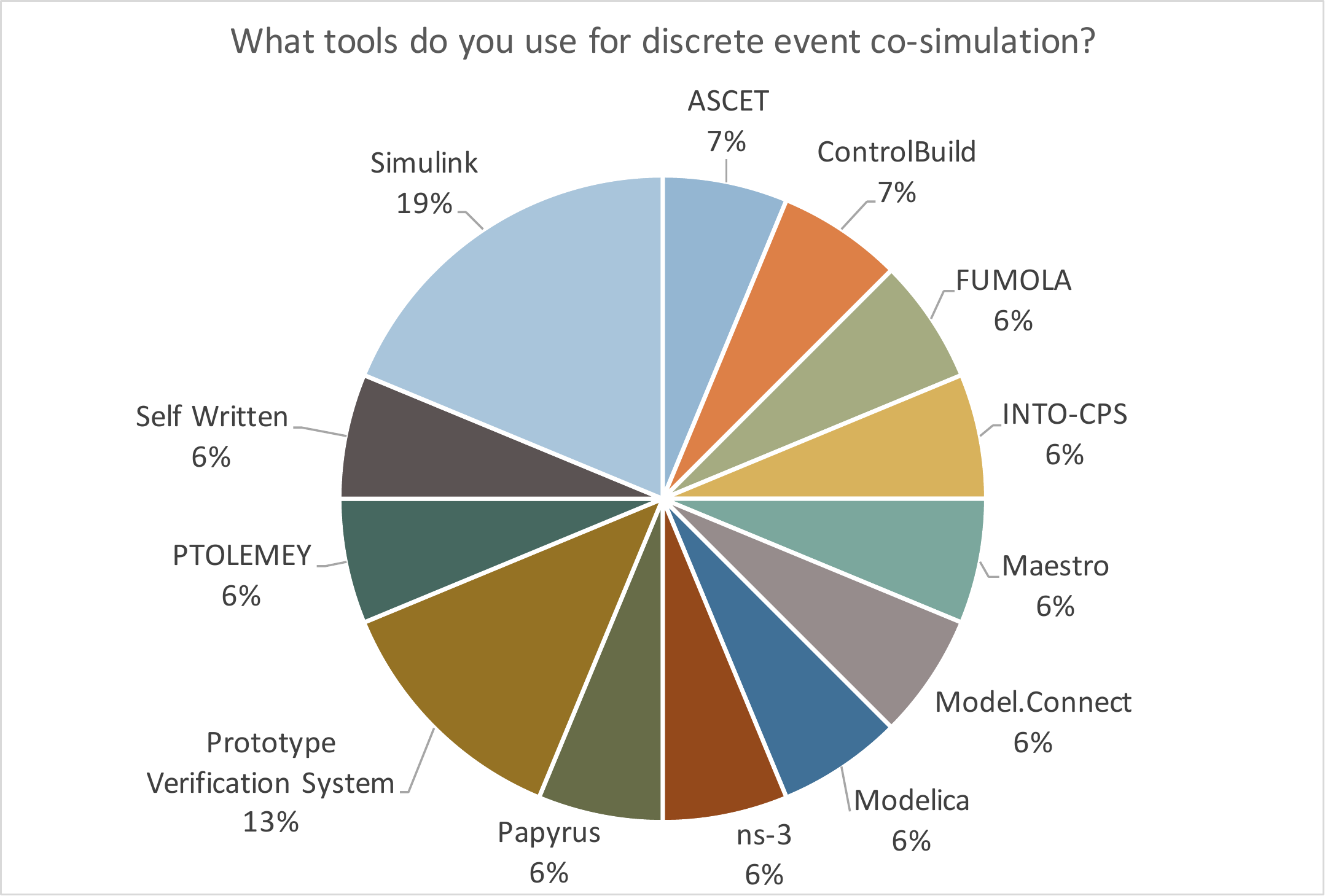}
\caption{Tools that experts use for discrete event co-simulation}
\label{fig:ToolsD}
\end{figure}

\newpage

\end{document}